\documentclass[preprintnumbers,groupedaddress,showpacs,showkeys]
{revtex4}
\usepackage{graphicx}

\def\sigb{\sigma^{\mbox{\tiny Born}}}
\def\siga{\sigma^\phi}
\def\Elab{E_{\mbox{\tiny lab}}}
\def\thetalab{\theta_{\mbox{\tiny lab}}}
\def\Emin{E_{\mbox{\tiny min}}}
\def\thetamin{\theta_{\mbox{\tiny min}}}
\def\thetamax{\theta_{\mbox{\tiny max}}}

\def\to{\rightarrow}
\def\lr{\leftrightarrow}

\def\lsim{\buildrel < \over {_\sim}}

\def\Ord{{\cal O}}

\def\spa#1.#2{\left\langle#1\,#2\right\rangle}
\def\spb#1.#2{\left[#1\,#2\right]}
\def\sect#1{section~{\ref{#1}}}

\def\fig#1{fig.~{\ref{#1}}}

\def\eqn#1{eq.~(\ref{#1})}
\def\eqns#1#2{eqs.~(\ref{#1}) and~(\ref{#2})}

\def\Eqns#1#2{Equations~(\ref{#1}) and~(\ref{#2})}

\begin{document}

\preprint{~~~~hep-ph/0402221\\}
\preprint{SLAC--PUB--10345}

\title{Radiative Corrections to the Azimuthal Asymmetry \\ 
in Transversely Polarized M\o{}ller Scattering\footnote{%
Research supported by the US Department of Energy under contract
DE-AC03-76SF00515.} }

\author{Lance Dixon}
\email[]{lance@slac.stanford.edu}
\author{Marc Schreiber}
\email[]{mschreib@slac.stanford.edu}
\affiliation{Stanford Linear Accelerator Center\\
Stanford University, Stanford, CA 94309, USA}


\begin{abstract}
Experiment E158 at SLAC can measure
an azimuthal asymmetry in single-spin, transversely polarized
M\o{}ller scattering, $e^{-\uparrow}e^- \to e^-e^-$, 
which arises from a QED rescattering phase.
We recompute the leading-order (one-loop) asymmetry,
confirming previous results, and calculate the 
leading logarithmic QED corrections due to initial-state radiation
from the beam and target electrons, and due to final-state radiation.
The size of these radiative corrections is quite sensitive to
experimental details, such as the acceptance in energy and in polar angle
of the scattered electron.  For E158, the corrections are modest,
increasing the parts-per-million asymmetry by roughly 1\%.
\end{abstract}


\maketitle


\section{Introduction \label{IntroSection}}

Single-spin triple product asymmetries, or asymmetries arising from 
transverse polarization, play a special role in scattering theory 
because they are directly sensitive to rescattering phases.  
An operator of the form 
$O \equiv {\bf S} \cdot ({\bf k} \times {\bf k}')$, 
where ${\bf S}$ is a spin and ${\bf k}$ and
${\bf k}'$ are two different particle momenta, is odd under a ``naive''
time-reversal operation which reverses all spins and momenta, but does not
exchange initial and final states.  A nonzero value for $O$ stems 
from terms in the covariant scattering amplitude
that are proportional to the Levi-Civita tensor
$\epsilon_{\mu\nu\rho\sigma}$, which always appears accompanied by
a factor of $i$.  Hence, in the absence of CP violation, 
a nonzero expectation value $\langle O \rangle$ requires an 
absorptive (imaginary) part for the amplitude, 
${\rm Im} \, T \neq 0$, which can be generated by rescattering, 
for example by one-loop diagrams containing intermediate 
two-particle cuts.

There have been many theoretical and experimental studies of single-spin
transversely-polarized asymmetries in a variety of contexts.
For instance, in the decay of a polarized neutron, 
$n^\uparrow \to p + e^- + \bar\nu_e$, 
an expectation value for ${\bf S}_n \cdot ({\bf k}_e \times {\bf k}_\nu)$
is produced by QED final-state interactions, which can therefore mask
truly ${\rm T}$-odd effects~\cite{JTW}. Analogous single-spin observables
in the decay of a polarized $Z$ boson to three hadronic jets, stemming
from QCD and electroweak final-state interactions, have been studied
theoretically~\cite{eezqqg} and bounded experimentally~\cite{SLDeezqqg}.
QCD final-state interactions can also play a role in generating azimuthal
single-spin asymmetries in semi-inclusive pion leptoproduction off
polarized protons at leading twist~\cite{BHS}.  
Similarly, a phase in the time-like electromagnetic proton form-factor 
from QCD final-state interactions can be detected by measuring 
transverse proton polarization in the reaction 
$e^+e^- \to p \bar{p}$~\cite{pformfactor}.

As a final example, QED rescattering phases produce
an azimuthal asymmetry in the elastic scattering
of electrons off transversely polarized protons, $ep^\uparrow \to ep$,
or transverse final-state polarization in the time-reversed reaction 
$ep \to ep^\uparrow$.  The QED asymmetry receives contributions not 
only from two-photon exchange with a single proton in the 
intermediate state, but also from inelastic hadronic intermediate 
states; the latter terms are difficult to compute directly, 
although they can be bounded experimentally~\cite{eptransverse}.

Perhaps the cleanest setting for studying such asymmetries is
in a process dominated by QED, such as transversely polarized
M\o{}ller scattering, $e^{-\uparrow}e^- \to e^-e^-$.  
Experiment E158 at SLAC performs M\o{}ller 
scattering of $\approx 45$ GeV polarized electrons off unpolarized 
target electrons at rest.  The prime goal of E158 is to measure the 
parity-violating right-left asymmetry in the cross section for 
longitudinal beam polarization, 
$A_{PV} = (\sigma_R-\sigma_L)/(\sigma_R+\sigma_L)$.  The right-left 
asymmetry is sensitive to $Z$ boson exchange, and potentially to 
new physics, such as a new $Z$ boson or contact interactions.  
The first measurement has yielded
$A_{PV} = (-175 \pm 30 ({\rm stat.}) \pm 20 ({\rm syst.})) 
\times 10^{-9}$~\cite{E158PV}.
While most of the E158 data were taken with the electron beam
polarized longitudinally in order to accomplish this measurement,
a fraction of the running was carried out with
transverse electron polarization, enabling
the measurement of an azimuthal asymmetry,
\begin{equation}
A_T(\phi) \equiv {2\pi \over \sigma^\uparrow + \sigma^\downarrow }
 { d (\sigma^\uparrow - \sigma^\downarrow) \over d\phi } 
\propto {\bf S}_e \cdot ({\bf k}_e \times {\bf k}'_e) 
\propto \sin\phi,
\end{equation}
where ${\bf S}_e$ is the spin of the incoming electron, with momentum
${\bf k}_e$, and $\phi$ is the azimuthal angle of the scattered electron
(with momentum ${\bf k}'_e$) around the beam direction, measured
from the direction of the transverse polarization.

\begin{figure*}
\includegraphics[width=9.5cm]{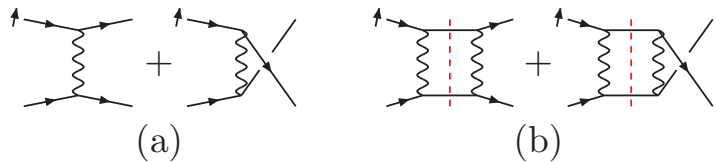}
\caption{\label{LOgraphs} (a) Tree-level graphs for electron-electron
scattering.  (b) One-loop graphs contributing to the azimuthal
asymmetry for transverse polarization.  Because an absorptive part
in the $s$ channel is required (the cut is indicated by the dashed line),
only box diagrams contribute.   The transverse spin of the beam electron
is indicated by the arrow next to that incoming line.}
\end{figure*}

In contrast to $A_{PV}$, a nonzero azimuthal asymmetry $A_T(\phi)$ 
can be generated by QED interactions alone.
The calculation of $A_T(\phi)$  for transversely polarized
M\o{}ller scattering at the leading one-loop order was performed by Barut
and Fronsdal in 1960~\cite{BarutFronsdal}, and by DeRaad and Ng in
1974~\cite{DeRaadNg}.  Because only the absorptive part of the scattering
amplitude contributes to this observable, an $s$-channel cut is required.
Hence only the box Feynman diagram enters, plus the version obtained by
exchanging the two identical outgoing electron legs, as depicted
in~\fig{LOgraphs}.  Besides the rescattering phase, the effect requires an
electron helicity flip.  For center-of-mass (CM) energies much larger than
the electron mass, $\sqrt{s} \gg m_e$, therefore, it takes the form
$A_T(\phi) = \alpha { m_e \over \sqrt{s} } f(\theta) \sin\phi$, where
$\alpha$ is the fine structure constant, $\phi$ and $\theta$ are
respectively the azimuthal and (CM frame) polar scattering angles, and $f$
is a function of $\theta$.

Since there are two identical electrons in the final state,
$f(\theta)$ must be odd under $\theta \to \pi - \theta$;
that is, a symmetric acceptance in $\theta$ (in an unsegmented detector)
will wash out the asymmetry in $\phi$.   
The E158 detector is well segmented in 
$\phi$ (twelve-fold), but coarsely segmented in $\theta$ (only
two-fold).  Fortuitously, the $\theta$ acceptance is almost
entirely in the backward hemisphere in the CM frame, 
leaving the sensitivity of E158 to $A_T(\phi)$ quite high.

The E158 CM energy is roughly 200~MeV, so the asymmetry is of order
$\alpha m_e/\sqrt{s} \sim 10^{-5}$.  This may seem small, but
it is two orders of magnitude larger than the electroweak asymmetry
$A_{PV}$.  Even though only a relatively small fraction of the data
was taken with transversely polarized electrons, a precision of the 
order of a few percent can be achieved for $A_T(\phi)$.  
One can either test QED at this level, or reverse
the logic and use the QED prediction as a detector calibration
or polarimeter~\cite{Woods}.

At the percent level of precision, it becomes important to investigate
the next-to-leading order (NLO), or $\Ord(\alpha^2)$, QED radiative corrections 
to $A_T(\phi)$. The full $\Ord(\alpha^2)$ calculation of the asymmetry
requires two-loop scattering amplitudes for $e^{-\uparrow}e^- \to e^- e^-$
and one-loop scattering amplitudes for $e^{-\uparrow}e^- \to e^- e^-\gamma$.
For $\sqrt{s} \gg m_e$, as in E158 kinematics, it would suffice to compute
these amplitudes in the limit where one takes $m_e \to 0$ after extracting
the leading $m_e/\sqrt{s}$ behavior from the diagrams.  This computation
should be feasible, because it is known how to perform all the relevant
two-loop four-point integrals~\cite{DoubleBoxes} and one-loop five-point
integrals~\cite{BDKPentagon} in this limit in dimensional regularization.
(Similar amplitudes without transverse polarization have already been
computed~\cite{SimilarAmps}.)

In the present paper, we calculate the largest of the $\Ord(\alpha^2)$
corrections, those that are enhanced by the large logarithm $\ln(s/m_e^2)$
due to collinear singularities in initial state radiation from both the
incoming beam electron and the target electron, as well as final-state
radiation.  The amplitude for $e^{-\uparrow}e^- \to e^- e^-\gamma$
factorizes in these collinear limits, so that its full kinematic
dependence is not required.  In an electron structure function
approach~\cite{ElectronSF}, analogues of the DGLAP splitting
kernels~\cite{DGLAP} enter our computation.  In the case of target
radiation and final-state radiation, only the unpolarized kernels are
required.  However, the radiation from the transversely polarized beam
electron also involves the analogues of kernels for the evolution of
transversely polarized quark distributions~\cite{AM}.  These
kernels can be obtained from the standard longitudinally polarized 
splitting amplitudes by a change of basis. 

We find that the magnitude of the leading-log NLO corrections 
is quite sensitive to the experimental cuts.  Initial state radiation
(ISR), for example, lowers the effective value of $s$, which could enhance
the asymmetry, since the leading-order asymmetry is proportional to 
$m_e/\sqrt{s}$. More importantly, ISR also skews the relation between 
polar angles in the post-radiation $e^-e^-$ CM frame and the lab frame,
changing the effective CM polar angle acceptance of the experiment.
Final state radiation (FSR) does not have either of these properties, and
typically produces smaller corrections to the asymmetry.

This paper is organized as follows.  In \sect{LOsection} we establish our
notation and review the leading-order azimuthal asymmetry 
prediction~\cite{BarutFronsdal,DeRaadNg}.  In \sect{NLOsection}
we describe the leading-log NLO corrections and present numerical results
for an experimental arrangement similar to E158.
In \sect{ConclusionSection} we present our conclusions.
In appendix~\ref{AppendixA} we give a derivation of the kernel
needed for evolution of the transversely polarized electron distribution.


\section{Notation and Leading-Order Results\label{LOsection}}

We consider the process
\begin{equation}
\label{kinematics}
e^{-\uparrow}(k_1)\ +\ e^-(k_2)\ \ \to\ \ e^-(k_1^\prime)\ +\ e^-(k_2^\prime)
\ \ [\ +\ \gamma(k_\gamma)\ ],
\end{equation}
where the photon is only present at next-to-leading order.
We use a right-handed $xyz$ coordinate system, writing momenta
$k^\mu = (k_t, k_x, k_y, k_z)$.  We take the energy of the beam 
electron in the lab frame to be $E$, and its momentum to be
in the $z$ direction: $k_1 = (E,0,0,\sqrt{E^2-m_e^2})$.
We let its polarization be in the negative $x$ direction~\footnote{%
An earlier version of this paper had an incorrect overall sign
for the leading-order asymmetry, which is remedied by choosing the
polarization to be in the {\it negative} (instead of positive)
$x$ direction.  Our results now agree with those of 
refs.~\cite{BarutFronsdal,DeRaadNg,Diaconescu}.
We thank Yury Kolomensky for pointing out this problem.}.
In the lab frame, the unpolarized target electron is at rest, 
$k_2 = (m_e,0,0,0)$.  The momentum of the detected scattered electron is 
$k_1^\prime = 
(\Elab, \sqrt{\Elab^2-m_e^2} \sin\thetalab \cos\phi,
        \sqrt{\Elab^2-m_e^2} \sin\thetalab \sin\phi,
        \sqrt{\Elab^2-m_e^2} \cos\thetalab)$;
its azimuthal angle $\phi$ increases from $0$ in the positive 
$x$ direction, through $\pi/2$ in the positive $y$ direction.

The Born-level differential cross section for M\o{}ller 
scattering, from the tree diagrams in \fig{LOgraphs}a, is
\begin{equation}
\label{sigLOexact}
\frac{d\sigb}{d\Omega} \bigg\vert_{\rm exact}
= \frac{\alpha^2}{2s} \cdot
\frac{(t^2+tu+u^2)^2 + 4m_e^2 (m_e^2 - t - u) (t^2 - tu + u^2) }{ t^2 u^2 } \,,
\end{equation}
where 
$s = (k_1+k_2)^2 = 2m_e (E+m_e)$, 
$u = (k_1^\prime - k_2)^2 = -2m_e (\Elab-m_e)$,
$t = (k_1^\prime - k_1)^2 = 4m_e^2 - s - u$.
(We include the statistical factor of 1/2 for identical electrons
in $d\sigma/d\Omega$, so such expressions should be integrated over
two-body phase-space for non-identical particles.)

The leading term in the cross section containing azimuthal dependence
arises at order $\alpha^3$, from the interference between the tree diagrams 
in \fig{LOgraphs}a and the box diagrams in \fig{LOgraphs}b.
The $\phi$-dependence at this order is given by~\cite{BarutFronsdal,DeRaadNg}
\begin{eqnarray}
\frac{d\siga}{d\Omega} \bigg\vert_{\rm exact}
 &=& - { \alpha^3 \over 8 } {m_e \over \sqrt{s}}
 \sin\theta \sin\phi \sqrt{1 - {4 m_e^2  \over s}} { 1 \over t^2 u^2 }
\nonumber \\ 
&&\hskip0.2cm
\times \biggl[
  3 (s - 4m_e^2) \biggl( t (u - s + 2m_e^2) \ln\Bigl( {-t \over s - 4m_e^2}\Bigr)
                 - u (t - s + 2m_e^2) \ln\Bigl( {-u \over s - 4m_e^2}\Bigr)
                 \biggr)
    - 2 (t-u) t u \biggr] \,.
\label{sigphiexact}
\end{eqnarray}
We have reproduced this result independently.

\Eqns{sigLOexact}{sigphiexact} include the exact dependence on the
electron mass.  However, in computing the NLO leading logarithms in 
$s/m_e^2$, we shall drop the terms suppressed by powers of $m_e^2/s$ 
in the leading-order asymmetry.  The error induced
by omitting these terms, for E158 kinematics, is much smaller than the 
size of the $\Ord(\alpha^2 \ln(s/m_e^2))$ corrections.
The Born-level and leading $\phi$-dependent cross sections then become,
\begin{equation}
\label{sigLO}
\frac{d\sigb}{d\Omega}
= \frac{\alpha^2}{2s} \cdot
\biggl( { t^2+tu+u^2 \over tu } \biggr)^2  \,,
\end{equation}
\begin{equation}
\frac{d\siga}{d\Omega}
 = - { \alpha^3 \over 8 } {m_e \over \sqrt{s}}
 \sin\theta \sin\phi\ { 1 \over t^2 u^2 }
\biggl[
  3 s \biggl( t (u - s) \ln\Bigl( {-t \over s}\Bigr)
            - u (t - s) \ln\Bigl( {-u \over s}\Bigr)
                 \biggr)
    - 2 (t-u) t u \biggr] \,,
\label{sigphi}
\end{equation}
and the kinematics can be simplified to
\begin{eqnarray}
&&
s = 2m_e E, \qquad
t = -2E\Elab(1-\cos\thetalab) = -{s\over2}(1-\cos\theta), \qquad
u = -2m_e\Elab = -{s\over2}(1+\cos\theta),
\label{LOstu}\\
&&
\Elab = {E\over2} (1 + \cos\theta), \qquad
\cos\thetalab = 1 - {m_e\over E} { 1-\cos\theta \over 1+\cos\theta },
\label{LOkinematics}
\end{eqnarray}
with $\theta$ the CM frame polar scattering angle.

\begin{figure*}
\includegraphics[width=9.5cm]{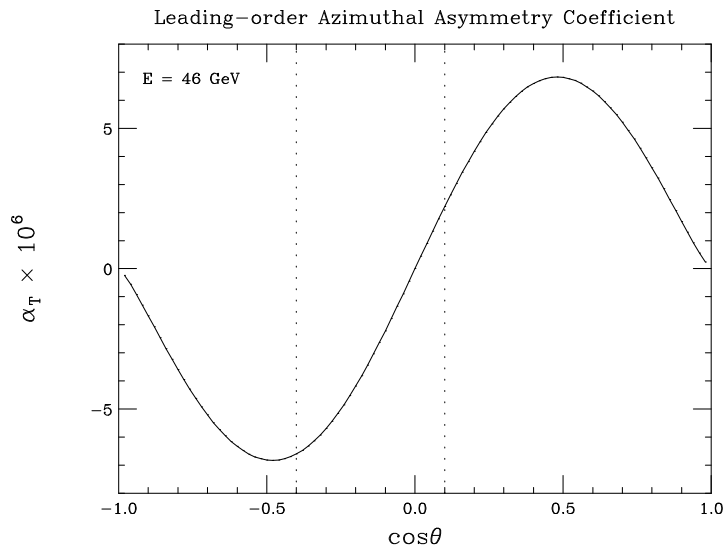}
\caption{\label{LOasymplot} The azimuthal asymmetry coefficient at leading
order, $\alpha_T^{LO}$, as a function of $\cos\theta$, for $E=46$ GeV,
$\sqrt{s} \approx 217$~MeV.  The vertical dotted lines indicate the 
approximate acceptance of E158 for leading-order kinematics.}
\end{figure*}

Writing the asymmetry as
\begin{equation}
A_T(\phi) \equiv {2\pi \over \sigma^\uparrow + \sigma^\downarrow }
 { d (\sigma^\uparrow - \sigma^\downarrow) \over d\phi } 
\equiv \alpha_T \cdot \sin\phi \,,
\label{asymdef}
\end{equation}
we have 
\begin{equation}
\alpha_T^{LO} = 
{1 \over \sin\phi} { d\siga/d\Omega \over d\sigb/d\Omega } \,.
\label{asymLO}
\end{equation}
In \fig{LOasymplot} the leading-order asymmetry coefficient,
$\alpha_T^{LO}$, is plotted as a function of CM polar angle, $\cos\theta$, 
for a beam energy of $E = 46$~GeV, or $\sqrt{s} \approx 217$~MeV.  
We set $\alpha = \alpha(\sqrt{s}) = 1/135.9$ here.  (E158 probes 
central scattering in the CM frame, with $|t|$ and $|u|$
ranging between $0.3s$ and $s$, so the difference between 
$\alpha(\sqrt{s})$ and 
$\alpha(\sqrt{|t|})$ or $\alpha(\sqrt{|u|})$ is negligible, 
less than 0.1\%.)  The asymmetry is of order parts per 
million at this energy.  

Note that $\alpha_T^{LO}$ is odd under $\theta \lr \pi-\theta$, 
or equivalently, that \eqn{sigphi} is odd under 
$t \lr u$.   This asymmetry is a consequence of having two identical electrons
in the final state.  In the CM frame, if one electron is at an angle
$(\theta,\phi)$, the other (at leading order) is at $(\pi-\theta,\phi+\pi)$.
Because $\sin\phi$ is odd under $\phi \lr \phi+\pi$, the coefficient
$\alpha_T^{LO}$ must be odd under $\theta \lr \pi-\theta$.
The odd behavior means that the integrated asymmetry seen by an experiment
integrating over a range in $\cos\theta$ is quite sensitive to the 
precise acceptance.  For example, a symmetric forward-backward acceptance 
in the CM frame leads to zero asymmetry at leading order.
The E158 polar-angle acceptance~\cite{E158PV}, 
$4.4~{\rm mrad} < \thetalab < 7.5~{\rm mrad}$,
corresponds mainly to the backward CM hemisphere for
leading-order kinematics, $-0.4 \lsim \cos\theta \lsim 0.1$,
as indicated by the dotted lines in \fig{LOasymplot}.

As mentioned in the introduction, the sensitivity to the acceptance 
could lead to relatively large QED corrections from hard photon radiation, 
which skews the kinematics of the $e^-e^- \to e^-e^-$ subprocess.
In the next section we investigate these corrections in more detail.


\section{NLO Calculation and Results \label{NLOsection}}

\begin{figure*}
\includegraphics[width=8.5cm]{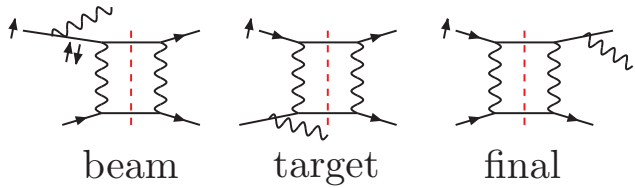}
\caption{\label{NLOgraphs} Diagrams contributing to the NLO
leading-log-enhanced corrections to the azimuthal asymmetry.
These graphs are to be interfered with corresponding graphs for
the Born process.  The exchange graphs are omitted.  Also shown, with
short arrows, are the transverse spin states of the initial electron, 
and of the quasi-on-shell electron line in the case of beam radiation.}
\end{figure*}

The leading-logarithmic QED corrections to the azimuthal asymmetry
arise from collinearly enhanced hard photon radiation.
These contributions can be divided into beam ($b$), target ($t$), and 
final-state ($f$) radiation, according to the electron line with 
which the photon is collinear, as shown in~\fig{NLOgraphs}.
In each of these limits, the $e^-e^- \to e^-e^-\gamma$ cross section 
factorizes into a collinear splitting probability~\cite{ElectronSF,DGLAP},
multiplied by the lower-order $e^-e^- \to e^-e^-$ cross section
evaluated for boosted kinematics.  In the construction of the asymmetry,
for the $\phi$-dependent numerator the boosted cross section is 
provided by $d\siga/d\Omega$ in~\eqn{sigphi}; for the denominator of 
the asymmetry it is $d\sigb/d\Omega$ in~\eqn{sigLO}.
We still have to pay a factor of $m_e/\sqrt{s}$ in $d\siga/d\Omega$; 
hence we can neglect powers of $m_e/\sqrt{s}$ in the splitting 
probabilities.

Although from the perspective of the lab frame one might not expect
radiation off of the target to be important, in the center of mass frame
such radiation is on an almost equal footing with radiation from the beam.
One difference, though, is that we have to track the transverse
polarization of the quasi-on-shell electron in the case of beam radiation,
as indicated by the opposing transverse arrows in~\fig{NLOgraphs}.  A
dilution of the transverse polarization will accompany the photon
radiation in this case.

We let $x$ denote the longitudinal momentum fraction retained
by an incoming or outgoing electron, after it has radiated a collinear
photon.  The $x\to 1$ limit represents emission of a soft photon.
In the leading-log approximation, we neglect the transverse momentum 
of the photon in computing the boosted kinematics of the 
$e^-e^- \to e^-e^-$ subprocess.
The integral over this small transverse momentum produces an
overall factor of $\ln(s/m_e^2)$.  
The unpolarized splitting probability for massless electrons is 
well known~\cite{ElectronSF}
\begin{equation}
P(x) = { 1 \over (1-x)_+ }  -  {1\over 2} (1+x) + {3\over4} \delta(1-x),
\label{unpolsplit}
\end{equation}
with the standard ``plus'' prescription definition,
\begin{equation}
\int_0^1 dx \frac{f(x)}{(1-x)_+} \equiv
\int_0^1 dx \frac{f(x)-f(1)}{1-x}.
\label{plusprescription}
\end{equation}
For the case of radiation from the transversely polarized beam,
we need to know the probability of a transverse spin flip.
This probability is unsuppressed in the massless electron limit,
because a transverse polarization state is a coherent superposition
of two different longitudinal (helicity) states.  Thus a helicity
flip is not required, only a different amplitude for the two different
electron helicity configurations, for a given photon helicity.
In appendix~\ref{AppendixA} we perform this computation.
The result can also be extracted from the QCD evolution equations for 
transversely polarized quarks~\cite{AM}, by converting color factors 
and coupling constants to the QED case: 
\begin{eqnarray}
P^{\uparrow\uparrow}(x) &=& { 1 \over (1-x)_+ }  
 -  {1\over4} (3+x) + {3\over4} \delta(1-x),
\label{noflipsplit} \\
P^{\uparrow\downarrow}(x) &=& {1\over4} (1-x),
\label{flipsplit}
\end{eqnarray}
where $P^{\uparrow\uparrow}$ ($P^{\uparrow\downarrow}$) is the 
splitting probability without (with) a transverse spin flip.
These probabilities satisfy 
$P^{\uparrow\uparrow}(x) + P^{\uparrow\downarrow}(x) = P(x)$.
It turns out that the $\delta(1-x)$ terms make a vanishing contribution 
to the azimuthal asymmetry, since they do not disrupt the leading-order 
kinematics, and the spin-flip probability vanishes in the soft limit
$x \to 1$.

In the case of ISR, because the radiated photon carries momentum, 
the effective CM energy-squared for the M\o{}ller scattering
decreases from $s = 2m_eE$ to $s^\prime = xs$.
In the case of FSR, the radiation happens after the scattering,
so $s^\prime = s$.  In radiative events, we use $\theta$
to denote the polar angle in the CM frame of the $e^-e^- \to e^-e^-$ 
subprocess.  To take into account experimental cuts, we need to
relate $\theta$ and $x$ to the lab variables $\thetalab$ and $\Elab$.
The relations are:
\begin{eqnarray}
s^\prime &=& xs, 
\qquad \Elab = x {E\over2}(1+\cos\theta),
\qquad \cos\thetalab
   = 1 - {m_e\over x \, E} { 1-\cos\theta \over 1+\cos\theta } \,,
\qquad \hbox{[beam]}
\label{beamkinematics}\\
s^\prime &=& xs, 
\qquad \Elab = {E\over2}(1+\cos\theta),
\qquad \cos\thetalab
   = 1 - {x \, m_e\over E} { 1-\cos\theta \over 1+\cos\theta } \,,
\qquad \hbox{[target]}
\label{targetkinematics}\\
s^\prime &=& s, 
\qquad \Elab = x {E\over2}(1+\cos\theta),
\qquad \cos\thetalab
   = 1 - {m_e\over E} { 1-\cos\theta \over 1+\cos\theta } \,.
\qquad \hbox{[final]}
\label{finalkinematics}
\end{eqnarray}
We define a model experimental acceptance in the lab frame by
\begin{equation}
A: \quad \Elab > \Emin,  \qquad \thetamin < \thetalab < \thetamax.
\label{expacceptance}
\end{equation}
Using eqs.~(\ref{beamkinematics}), (\ref{targetkinematics}) and 
(\ref{finalkinematics}) the acceptance $A$ can be translated into 
acceptances $A_b$, $A_t$, and $A_f$ bounding the integration region for 
$x$ and $\theta$ in the respective correction terms 
(and also the region $A_0$ which bounds the $\theta$ integral for 
leading-order, non-radiative events).

Including collinear radiation, the relevant terms in the differential
cross section are modified as follows,
\begin{eqnarray}
{d\sigb(s) \over d\Omega} &\to& 
{d\sigb(s) \over d\Omega}
+ {\alpha\over\pi} \ln\Bigl({s\over m_e^2}\Bigr) 
\sum_{i=b,t,f} \int_0^1 dx 
 \, P_i^{\mbox{\tiny Born}}(x) {d\sigb(s) \over d\Omega} \,,
\label{sigbwithrad} \\
{d\siga(s) \over d\Omega} &\to& 
{d\siga(s) \over d\Omega}
+ {\alpha\over\pi} \ln\Bigl({s\over m_e^2}\Bigr) 
\sum_{i=b,t,f} \int_0^1 dx 
  \, P_i^\phi(x) {d\siga(s) \over d\Omega} \,,
\label{sigawithrad}
\end{eqnarray}
where $P_b^\phi = P^{\uparrow\uparrow}(x) - P^{\uparrow\downarrow}(x)$,
$P_b^{\mbox{\tiny Born}} = P_t^{\mbox{\tiny Born}} = P_f^{\mbox{\tiny Born}} 
 = P_t^\phi = P_f^\phi = P(x)$.
We insert \eqns{sigbwithrad}{sigawithrad} into \eqn{asymLO},
perform the integrals over the respective acceptances in both the numerator
and denominator of the asymmetry, expand the result in 
$\alpha$, and thus obtain for the leading-log-corrected asymmetry 
coefficient, 
\begin{equation}
\alpha_T^{LL} = \alpha_T^{LO} ( 1 + \delta_b + \delta_t + \delta_f ),
\label{alphaTdecomp}
\end{equation}
where
\begin{eqnarray}
\alpha_T^{LO} &=& { N_0 \over D_0 } \,,
\label{alphaTLO} \\
\delta_i &=&
{\alpha(\sqrt{s}) \over \pi} \ln\Bigl( {s \over m_e^2} \Bigr) 
\biggl[ { N_i \over N_0 } - { D_i \over D_0 } \biggr] \,,   
\qquad i = b,t,f.
\label{deltadef}
\end{eqnarray}
Here the leading-order integrated results are 
\begin{equation}
 N_0 = \int_{A_0} d\cos\theta \, {d\siga(s) \over d\Omega}, 
\qquad
 D_0 = \int_{A_0} d\cos\theta \, {d\sigb(s) \over d\Omega} \,.
\label{D0N0}
\end{equation}
The radiative terms are
\begin{eqnarray}
 N_b &=& \int_{A_b} dx \, d\cos\theta \, 
 [ P^{\uparrow\uparrow}(x) - P^{\uparrow\downarrow}(x) ]
  \, {d\siga(xs) \over d\Omega},
\qquad
 D_b = \int_{A_b} dx \, d\cos\theta \, P(x) \, {d\sigb(xs) \over d\Omega} \,,
\label{DbNb}\\
 N_t &=& \int_{A_t} dx \, d\cos\theta \, P(x) \, {d\siga(xs) \over d\Omega},
\qquad
 D_t = \int_{A_t} dx \, d\cos\theta \, P(x) \, {d\sigb(xs) \over d\Omega}\,,
\label{DtNt}\\
 N_f &=& \int_{A_f} dx \, d\cos\theta \, P(x) \, {d\siga(s) \over d\Omega},
\qquad
 D_f = \int_{A_f} dx \, d\cos\theta \, P(x) \, {d\sigb(s) \over d\Omega}\,.
\label{DfNf}
\end{eqnarray}

In table~\ref{Emintable} we present results for the azimuthal asymmetry
coefficient for $E = 46$~GeV, as a function of the minimum accepted 
energy $\Emin$, for $4.4 < \thetalab < 7.5$~mrad.  We give the
leading-order result integrated over the acceptance, $\alpha_T^{LO}$;
the beam, target and final-state fractional corrections $\delta_b$,
$\delta_t$ and $\delta_f$; and the QED-corrected result $\alpha_T^{LL}$.
The leading-order result does not depend on $\Emin$ until 
$\Emin > 13$~GeV; at that point the $\Emin$ cut starts to
remove the most backward-scattered electrons (in the CM frame),
which have the lowest energies in the lab frame.  The corrections
from beam and target radiation have opposite sign, because
such radiated photons skew in opposite directions the relation between 
the subprocess CM frame and the lab frame, as indicated
by~\eqns{beamkinematics}{targetkinematics}.
For beam radiation, as $x$ decreases from 1,
a given angle in the subprocess CM frame boosts to a
larger angle in the laboratory frame.  Hence,
for small $x$, the experimental cuts now sample some of the
CM forward hemisphere, where the LO asymmetry is positive.
Thus $\delta_b$ is negative.
For target radiation, however, as $x$ decreases from 1,
the boost back to the lab frame becomes larger and
the resulting CM angles boost to smaller lab frame angles.
Now small $x$ forces the experimental
cuts to sample more of the CM backward hemisphere, where the LO
asymmetry can be even more negative.  Thus $\delta_t$ is positive.
It is also larger in magnitude than $\delta_b$, which may be 
due to the depolarization of the beam by ISR:  
$P^{\uparrow\uparrow}(x) - P^{\uparrow\downarrow}(x) < P(x)$.
As $\Emin$ decreases, both $\delta_b$ and $\delta_t$ increase in
magnitude, as more hard radiative events are permitted, which
skew the kinematics more.  Final state radiation does not alter the 
LO relation between $\theta$ and $\thetalab$.  It only has an 
effect via the minimum energy cut, which affects the effective
$\cos\theta$ acceptance through \eqn{finalkinematics} for $\Elab$.
Indeed, $\delta_f$ decreases as $\Emin$ is lowered.

\begin{table}
\caption{\label{Emintable} Azimuthal asymmetry coefficient as
a function of $\Emin$ for $E=46$~GeV, $\thetamin=4.4$\,mrad, 
$\thetamax=7.5$\,mrad.}
\begin{ruledtabular}
\begin{tabular}{c|ccccc}
$\Emin$~(GeV) & $\alpha_T^{LO}\times 10^6$
            & $\delta_b$ & $\delta_t$ & $\delta_f$ 
            & $\alpha_T^{LL}\times 10^6$
\\ \hline
\hskip5pt   8 &  -3.7949  &  -0.0221   &  0.0452  &  0.0011  &  -3.8826 \\
\hskip5pt  10 &  -3.7949  &  -0.0121   &  0.0282  &  0.0015  &  -3.8562 \\
\hskip5pt  12 &  -3.7949  &  -0.0060   &  0.0165  &  0.0019  &  -3.8348 \\
\hskip5pt  14 &  -3.4180  &  -0.0040   &  0.0109  &  0.0022  &  -3.4414 \\
\end{tabular}
\end{ruledtabular}
\end{table}

Table~\ref{thetatable} presents azimuthal asymmetry results with
the minimum accepted energy $\Emin$ held fixed at 13~GeV, and
the minimum angle fixed at $\thetamin = 4.4$\,mrad, but varying 
the maximum angle $\thetamax$.  Now the variation in the QED-corrected
result is dominated by the variation in the leading-order term
$\alpha_T^{LO}$, since the leading-order acceptance is changing.  
However, the size of $\delta_b$ and $\delta_t$ also depends strongly
on $\thetamax$, presumably because the {\it slope} of the leading order
asymmetry at $\theta = \thetamax$ (the left dotted line in
\fig{LOasymplot}) is also changing; the slope determines how effective
the skewed kinematics are in altering the asymmetry.

\begin{table}
\caption{\label{thetatable} Azimuthal asymmetry coefficient as
a function of $\thetamax$ for $E=46$~GeV, $\Emin=13$~GeV, 
$\thetamin=4.4$\,mrad.}
\begin{ruledtabular}
\begin{tabular}{c|ccccc}
$\thetamax$~(mrad) & $\alpha_T^{LO}\times 10^6$
            & $\delta_b$ & $\delta_t$ & $\delta_f$ 
            & $\alpha_T^{LL}\times 10^6$
\\ \hline
\hskip5pt  6.5 &  -2.6358  &  -0.0102   &  0.0241  &  0.0016  &  -2.6724 \\
\hskip5pt  7.0 &  -3.2762  &  -0.0061   &  0.0166  &  0.0019  &  -3.3103 \\
\hskip5pt  7.5 &  -3.7949  &  -0.0044   &  0.0121  &  0.0021  &  -3.8241 \\
\hskip5pt  8.0 &  -3.8039  &  -0.0043   &  0.0120  &  0.0021  &  -3.8330 \\
\end{tabular}
\end{ruledtabular}
\end{table}


\section{Conclusions and Outlook \label{ConclusionSection}}

In this paper we computed the leading-logarithmic QED corrections 
to the azimuthal symmetry in transversely polarized M\o{}ller scattering,
which relies on a one-loop rescattering phase, and is currently being 
measured by the E158 experiment. The correction term arising from  
radiation off the beam electron involves a transverse spin-flip
splitting probability analogous to that encountered in the
QCD evolution of transversely polarized quark distributions,
which dilutes the beam polarization.  The corrections from radiation 
off the beam and target are opposite in sign, because they skew the 
kinematic relation between the subprocess-center-of-mass frame and 
lab frame in opposite directions.  Final-state radiation is smaller
in size.  The net effect depends on the cuts, but is typically
about a 1\% increase in the magnitude of the asymmetry.
This shift is somewhat below the anticipated precision of E158s
measurement of a few percent.  In principle, therefore, the present 
QED prediction, combined with the E158 measurement, could be used as 
an alternate way to measure the beam polarization, or calibrate the
azimuthal response of the detector.   Finally, the computation
of the non-logarithmically enhanced QED corrections is a feasible
future project, though probably not mandated by the presently 
achievable experimental precision.


\begin{acknowledgments}
We thank Yury Kolomensky, Krishna Kumar and Mike Woods for suggesting this
problem and for helpful discussions and information about the E158
experiment.  We are also grateful to Stan Brodsky and Michael Peskin
for useful conversations and comments on the manuscript.
\end{acknowledgments}


\appendix

\section{Evolution of transverse electron polarization \label{AppendixA}}

Collinear photon radiation can produce a transverse spin flip for a 
massless electron because the transverse spin state is a coherent 
superposition of both longitudinal spin (helicity) states.
There is no longitudinal spin flip in the massless limit, but for a 
given photon helicity, the amplitude for radiation depends on the 
electron helicity.  In the transverse basis, this dependence
generates the spin flip.  

Explicitly, the transverse states $|\uparrow\rangle$ and
$|\downarrow\rangle$ are given in terms of longitudinal states 
$|+\rangle$ and  $|-\rangle$ by
\begin{equation}
|\uparrow\rangle = {1\over \sqrt{2}} ( |+\rangle + |-\rangle ),
\qquad
|\downarrow\rangle = {1\over \sqrt{2}} ( |+\rangle - |-\rangle ).
\label{transversestates}
\end{equation}
The $x$-dependence of the amplitudes for collinear splitting,
$e \to e\gamma$, in the helicity basis can be extracted from
analogous results for the $q \to qg$ splitting amplitudes in QCD
(see {\it e.g.} ref.~\cite{CollLimits}).  The non-vanishing,
helicity-conserving amplitudes are,
\begin{eqnarray}
 {\cal A}(e^{(+)} \to e^{(+)} \, \gamma^{(+)}) &=&
 {\cal A}(e^{(-)} \to e^{(-)} \, \gamma^{(-)}) = { 1 \over \sqrt{1-x} } \,,
\label{ptopp} \\
 {\cal A}(e^{(-)} \to e^{(-)} \, \gamma^{(+)}) &=& 
 {\cal A}(e^{(+)} \to e^{(+)} \, \gamma^{(-)}) = { x \over \sqrt{1-x} } \,.
\label{mtomp}
\end{eqnarray}
The $x$-dependence of the usual unpolarized splitting probability 
for $x<1$, $P(x) \propto (1+x^2)/(1-x)$, can easily be recovered 
by summing the squares of these amplitudes.  Here we wish to transform
these amplitudes to the transverse electron spin 
basis~(\ref{transversestates}), 
\begin{eqnarray}
 {\cal A}(e^{\uparrow} \to e^{\uparrow} \, \gamma^{(+)}) &=&
 {1\over\sqrt{2}} \left(\matrix{ 1 & 1 \cr } \right) 
  \left( \matrix{ {1\over \sqrt{1-x}} & 0 \cr 
                    0 &  {x\over \sqrt{1-x}} \cr } \right)
 {1\over\sqrt{2}} \left(\matrix{ 1 \cr 1 \cr } \right) 
   = { 1 + x \over 2 \sqrt{1-x} } \,,
\label{uptoup} \\
 {\cal A}(e^{\uparrow} \to e^{\downarrow} \, \gamma^{(+)}) &=&
 {1\over\sqrt{2}} \left(\matrix{ 1 & 1 \cr } \right) 
  \left( \matrix{ {1\over \sqrt{1-x}} & 0 \cr 
                    0 &  {x\over \sqrt{1-x}} \cr } \right)
 {1\over\sqrt{2}} \left(\matrix{ 1 \cr -1 \cr } \right) 
   = { \sqrt{1-x} \over 2} \,.
\label{uptodown}
\end{eqnarray}
The amplitudes for the case of negative photon helicity have the 
same magnitudes, using parity.  Note that the relative phase of 
the amplitudes given in \eqns{ptopp}{mtomp} is important in
\eqns{uptoup}{uptodown}; it can be fixed by requiring that the 
amplitudes become independent of the electron helicity in the 
soft photon limit $x \to 1$.

The square of \eqn{uptodown} gives the $x$-dependence of the 
transverse spin-flip splitting probability in \eqn{flipsplit}, 
$P^{\uparrow\downarrow}(x) = {1\over4} (1-x)$.  This term
needs no plus-prescription regularization as $x\to1$; nor
is there a $\delta(1-x)$ term.  The square of \eqn{uptoup} 
gives the $x$-dependence of $P^{\uparrow\uparrow}(x)$ in 
\eqn{noflipsplit}; here plus-prescription regularization is required.
The overall normalization of $P^{\uparrow\uparrow}$ and 
$P^{\uparrow\downarrow}$ can be fixed by requiring their sum
to be equal to $P(x)$ in \eqn{unpolsplit}.
The $\delta(1-x)$ term in $P^{\uparrow\uparrow}$ can be inferred 
from electron number conservation, 
$\int_0^1 dx \, P(x) 
= \int_0^1 dx \, [ P^{\uparrow\uparrow}(x) + P^{\uparrow\downarrow}(x) ]
= 1$.


\end{document}